\documentclass[a4paper,fleqn]{cas-dc}
\usepackage[numbers,sort&compress]{natbib}

\def\tsc#1{\csdef{#1}{\textsc{\lowercase{#1}}\xspace}}
\tsc{WGM}
\tsc{QE}

\usepackage[dvipsnames]{xcolor}
\usepackage[most]{tcolorbox}
\usepackage{listings}
\usepackage{caption}
\usepackage{amsmath}
\usepackage{float}

\restylefloat{table}

\definecolor{codegreen}{rgb}{0,0.6,0}
\definecolor{codegray}{rgb}{0.5,0.5,0.5}
\definecolor{codepurple}{rgb}{0.58,0,0.82}
\definecolor{backcolour}{rgb}{0.95,0.95,0.92}
\definecolor{codeorange}{rgb}{1,0.49,0}
\definecolor{blue}{rgb}{0.01, 0.61, 0.98}

\lstdefinestyle{mystyle}{
    backgroundcolor=\color{backcolour},
    commentstyle=\color{codegreen},
    keywordstyle=\color{magenta},
    numberstyle=\tiny\color{codegray},
    stringstyle=\color{codepurple},
    basicstyle=\footnotesize\ttfamily,
    breakatwhitespace=false,
    breaklines=true,
    captionpos=b,
    keepspaces=true,
    numbers=left,
    numbersep=8pt,
    showspaces=false,
    showstringspaces=false,
    showtabs=false,
    tabsize=2,
    frame=single,  
    rulecolor=\color{black},  
    linewidth=0.9\linewidth, 
    xleftmargin=0.05\linewidth 
}

\begin{document}

\shortauthors{Spadon et~al.}

\shorttitle{Maritime Tracking Data Analysis and Integration with \textsc{AISdb}}
\title [mode = title]{Maritime Tracking Data Analysis and Integration with \textsc{AISdb}}

\tnotemark[1]

\tnotetext[1]{\textsc{AISdb} is a contribution from the Canadian Foundation for Innovation's \textsc{MERIDIAN} Cyberinfrastructure, hosted at the Institute for Big Data Analytics (\textsc{IBDA}) from Dalhousie University in Halifax, NS --- Canada.}
   
\author[1,2]{Gabriel Spadon}[orcid=0000-0001-8437-4349]
\ead{spadon@dal.ca}
\cormark[1]

\author[1,2]{Jay Kumar}[orcid=0000-0003-4915-9701]
\ead{jay.kumar@dal.ca}

\author[1,2]{Jinkun Chen}[orcid=0009-0004-5792-2097]
\ead{jinkun.chen@dal.ca}

\author[2]{Matthew Smith}
\ead{matthew.smith@dal.ca}

\author[2]{Casey Hilliard}[orcid=0000-0001-5890-6006]
\ead{casey.hilliard@dal.ca}

\author[2]{Sarah Vela}[orcid=0000-0002-3007-7469]
\ead{svela@dal.ca}

\author[2]{Romina Gehrmann}[orcid=0000-0002-3099-2771]
\ead{rgehrmann@dal.ca}

\author[3]{Claudio DiBacco}[orcid=0000-0002-9662-5798]
\ead{claudio.dibacco@dfo-mpo.gc.ca}

\author[2]{Stan Matwin}[orcid=0000-0001-6629-8434]
\ead{stan@cs.dal.ca}

\author[1]{Ronald Pelot}[orcid=0000-0001-8837-0409]
\ead{ronald.pelot@dal.ca}

\affiliation[1]{organization={Department of Industrial Engineering, Dalhousie University}, city={Halifax, NS}, country={Canada}}
\affiliation[2]{organization={Faculty of Computer Science, Dalhousie University}, city={Halifax, NS}, country={Canada}}
\affiliation[3]{organization={Fisheries and Oceans Canada}, city={Dartmouth, NS}, country={Canada}}

\cortext[cor1]{~\textit{Corresponding author}}

\begin{abstract}
Efficiently handling Automatic Identification System (AIS) data is vital for enhancing maritime safety and navigation, yet is hindered by the system's high volume and error-prone datasets. This paper introduces the Automatic Identification System Database (AISdb), a novel tool designed to address the challenges of processing and analyzing AIS data. AISdb is a comprehensive, open-source platform that enables the integration of AIS data with environmental datasets, thus enriching analyses of vessel movements and their environmental impacts. By facilitating AIS data collection, cleaning, and spatio-temporal querying, AISdb significantly advances AIS data research. Utilizing AIS data from various sources, AISdb demonstrates improved handling and analysis of vessel information, contributing to enhancing maritime safety, security, and environmental sustainability efforts.
\end{abstract}




\maketitle

\section{Introduction and Rationale}
\label{sec:intro}

The Automatic Identification System (AIS) utilizes radio technology, enabling vessels to transmit encoded messages from terrestrial and satellite receivers~\cite{AIS, IMO}. AIS was initially conceived to track vessels and prevent collisions at sea but has evolved significantly, expanding its scope of applications to encompass diverse fields such as traffic surveillance, trends analysis, vessel security, and monitoring environmental pollution~\cite{peng2022establishment, yang2019big, fournier2018past}. The content of the broadcast messages may include real-time positional data, additional vessel characteristics, and voyage details, thereby increasing awareness within transportation networks, i.e., decreasing the risk of vessel collisions~\cite{Goerlandt2011:ShipCollision, Chen2019:ShipCollision}. The AIS data falls under the category of spatiotemporal data, describing multiple aspects of an object that evolve with time. Given the high frequency of AIS message transmissions by vessels~\cite{yang2019big, harati2007automatic}, the resulting data stream comprises millions of messages daily~\cite{millefiori2021covid}. Additionally to the challenge of dealing with such a vast volume of data, AIS often contains errors~\cite{fu2017finding, li2018spatio}, ranging from damaged to duplicated or inaccurate messages, attributable to human error, equipment malfunction and connectivity issues. Consequently, ensuring the quality of AIS data is imperative for real-time applications~\cite{haranwala2023data, mustafa2021gtraclus, rong2020data}.

With the aid of AIS data, researchers and industry professionals can leverage the AIS data knowledge while training machine learning models for various purposes, including predictive modeling of vessel trajectories~\cite{nguyen2018multi, patmanidis2016maritime, zhang2018ais, uney2019data}, anomaly detection in navigational patterns, and environmental impact assessments.
Predictive modeling of vessel trajectories leverages historical AIS data to forecast future vessel movements, essential for optimizing shipping routes, reducing fuel consumption, and preventing collisions~\cite{rong2020data, spadon2022unfolding, newaliya2021review, le2013unsupervised, d2018maritime, forti2019unsupervised}.
Anomaly detection in navigational patterns helps identify unusual or potentially dangerous behavior in maritime traffic, enhancing security and safety measures~\cite{campbell2022detection, forti2019anomaly}.
Environmental impact assessments utilize AIS data to study the effects of maritime activities on marine ecosystems, aiding in the development of sustainable practices and policies~\cite{peng2022establishment, pichegru2022maritime}.
Additionally, AIS data combined with machine learning can be used for traffic density analysis, which helps understand and manage port congestion and develop intelligent maritime traffic systems~\cite{faghih2014accident, abualhaol2018mining}.
Besides, machine learning models trained on AIS data can improve search and rescue operations by predicting the drift patterns of vessels or objects lost at sea~\cite{varlamis2018detecting, zhou2020performance, chen2019study}.
Moreover, these models can assist fisheries management by tracking and analyzing fishing vessel activities to ensure regulation compliance~\cite{ferreira2022semi}.

Several Python packages and approaches have been made available for researchers to preprocess the positional (trajectory or kinematics) data, such as PyVT~\cite{DBLP:journals/softx/LiRL23}, PyMove~\cite{arina2019}, and scikit-move~\cite{DBLP:journals/jstatsoft/PappalardoSBP22}. PyVT loads the data from a CSV file and facilitates vessel trajectory data preprocessing, trajectory clustering, and trajectory visualization. Similarly, PyMove and scikit-move can perform interpolation, outlier removal, filtering, and data mining of trajectory while extending the existing Pandas and scikit-learn library. However, these tools are not designed to use the available processing resources for big data processing.
PTRAIL~\cite{DBLP:journals/softx/HaidriHBRFS22} uses NumPy and enables parallel processing for functions like filtering, linear and cubic interpolation, and feature extraction (e.g., day of the week, traveled distance, acceleration, bearing). TraClets~\cite{DBLP:journals/softx/KontopoulosMT23} uses kinematic data into image representations of trajectories for the classification task. ADP~\cite{TANG2021109041} implements a trajectory compression approach to reduce the size of trajectories.
However, these techniques cannot cope with extracting vessel characteristics and voyage information, real-time data ingestion, efficient big data management, and seamless data integration.

To solve the aforementioned issues, we have developed a tool called the Automatic Identification System Database (AISdb)\footnote{~See https://aisdb.\-meridian.cs.dal.ca/doc.}. Our objective is to allow researchers to ensure reproducibility in future AIS-based research while dealing with erroneous and noisy messages from AIS data. AISdb is a Python 3 library that allows users to: {1 --} load the data either from a real-time stream, encoded NM4, or CSV file into the database, {2 --} remove noise, {3 --} cure data by adding missing vessel information,{4 --} interpolate and segment vessel trajectories, {5 --} filter data concerning time and geographical frame, {6 --} access interactive trajectory visualization, {7 --} integrate environmental data, {8 --} transform data, and {9 --} export data. For each function, AISdb provides detailed examples and sample data to illustrate the use of the library. Tools like PyVT, PTRAIL, TraClets, and ADP can be combined with AISdb to produce data pipelines that can be used on machine learning vessel modeling tasks.

AISdb is a versatile solution for storing data on local machines using SQLite and servers using PostgreSQL databases, making it well-suited for complex and collaborative environments. AISdb enforces specific database constraints to optimize the structuring and indexing of large spatial AIS datasets and to eliminate irrelevant AIS messages resulting from transmission noise. Moreover, it transforms AIS data into valuable insights by accurately capturing vessel movements for research and industrial purposes. AISdb also seamlessly integrates with various environmental datasets, enabling analyses of maritime traffic and its impact on marine ecosystems. As a result, AISdb has become a tool providing academics and industry professionals with a platform for studying the marine environment and ship mobility.

To outline AISdb contributions this paper is structured as follows: Section~\ref{sec:history} provides an overview of AISdb, its origins, and the development team; Section~\ref{sec:arch} describes the architecture of AISdb, outlining the functionality of each module; Section~\ref{sec:sci} highlights the relevance of AISdb in research initiatives; and, Section~\ref{sec:conclusions} presents the final remarks.

\section{\textsc{AISdb} History}
\label{sec:history}

AISdb results from over ten years of research on AIS data carried out by the Institute of Big Data Analytics (IBDA) at the Faculty of Computer Science from Dalhousie University. The codes generated by these researches were organized and centralized in an internal repository to facilitate further AIS-based research and partnerships. With increasing interest and cross-disciplinary collaboration at Dalhousie University, AISdb began to take shape to organize and centralize data. It was made an open-source repository that was widely disseminated among researchers and industry partners.

The Marine Environmental Research Infrastructure for Data Integration and Application Network (MERIDIAN), hosted at the IBDA, led the development of the AISdb and collaborated with local partners to incorporate their needs into the AISdb platform. During this process, the Maritime Risk and Safety Research Group (MARS) at the Department of Industrial Engineering from the Faculty of Engineering at Dalhousie University collaborated with MERIDIAN and became a key platform user. MARS utilized AISdb to conduct research on search and rescue in Arctic waters. This partnership between MARS and MERIDIAN has been instrumental in supporting the research and development of AISdb, with the aim to push the boundaries of AIS research.

MARS and MERIDIAN, in collaboration with the Department of Fisheries and Oceans (DFO) of Canada, have secured support to continue enhancing AISdb. This initiative aims to integrate advanced machine learning techniques to improve the analysis and interpretation of AIS data, such as through vessel modeling and forecasting. The goal is to broaden awareness and display the practical applications of AIS data utilization among local communities in Atlantic Canada, fostering better understanding and engagement with maritime activities and data-driven decision-making.

\section{Software Description}
\label{sec:arch}

AISdb is an advanced Python library that handles and analyzes AIS data effectively. The main goal of this package is to provide efficient processing and visualization of spatiotemporal AIS data, which can be particularly useful for data science analysis and machine learning applications.

\begin{figure}[!b]
    \includegraphics[width=8cm]{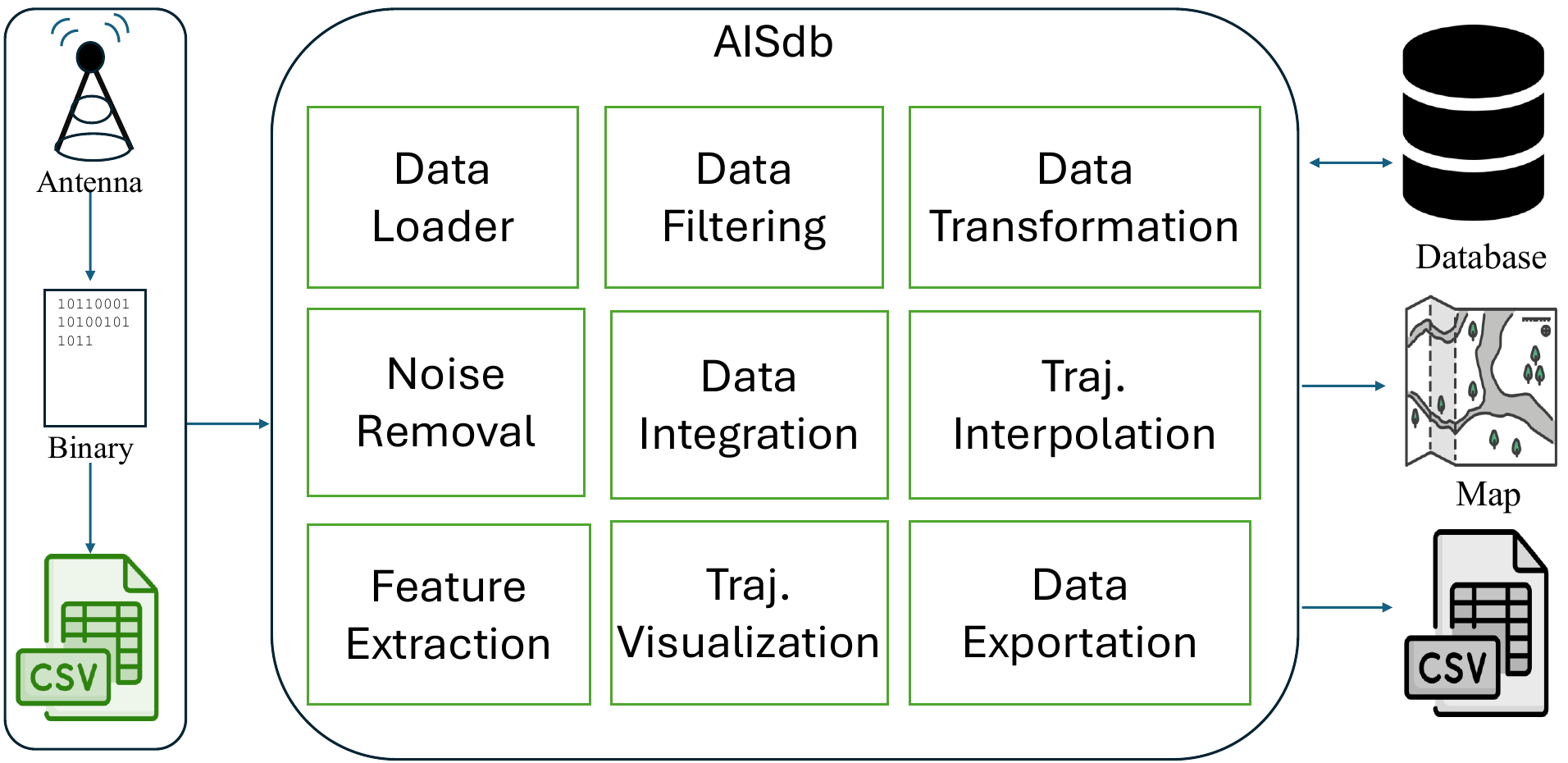}
    \centering
    \caption{The technical architecture of AISdb workflow.}
    \label{fig:architecture}
\end{figure}

\subsection{Software Architecture}

The process of processing AIS data is shown in Figure~\ref{fig:architecture}.
The AISdb package enables users to input data directly from an online stream from an antenna receiver, an encoded binary file, and CSV files. The data are loaded and stored in the user's selected database. Subsequently, we provide a detailed view of the package modules and functionalities.

\subsection{Software Functionalities}
\subsubsection{Decoding AIS Messages}

The decoder module is one of the main components of AISdb.
It plays a crucial role in managing the decoding and decapsulation of AIS messages. This module boasts essential features that ensure data integrity and simplify further analysis of AIS data.
One of the key functions of this module is its ability to convert raw, binary-encoded AIS messages into a structured, easily understandable format. This data can originate from various sources, including compressed zip files, binary-encoded messages in NM4 format, or even live-feed AIS data from a self-hosted AIS antenna that runs on a micro-controller subsystem and communicates directly with the AISdb live-feed API. Additionally, the decoder module features a generator capable of parsing continuous streams of AIS messages into a format ready for immediate analysis. This feature is handy for real-time data processing applications, supporting dynamic maritime monitoring and decision-making processes. Specifically, it can be utilized as a data loader to train neural networks to perform specific tasks based on the knowledge extracted from historical data.

The live-stream support in AISdb was developed to utilize live data feeds through two strategically placed antennas: one at the Dalhousie University campus and another at the Marine Research Station in Sandy Cove, Nova Scotia, Canada, in collaboration with the Coastal Environmental Observation Technology and Research group. The antennas cover regional data from Halifax Harbour to about 5 km offshore and are shared freely with partners and AISdb users.

\subsubsection{Data Loader}

The data loader module stores the parsed raw data in the database. The database connectivity module is the foundation of AISdb's functionality, providing essential functions for establishing and managing database connections. This module features the ConnectionType enumeration, which classifies the supported database types as SQLite and PostgreSQL. This classification is significant since it enables the dynamic selection of database interaction strategies, allowing AISdb to adapt seamlessly to various storage requirements and environments.
The DBConn class, the cornerstone of this module, provides a comprehensive API for initiating connections. This API standardizes query execution and data handling procedures across different database systems, ensuring a consistent user experience regardless of the database backend. To cater to specific database systems, AISdb includes two subclasses based on the supported databases: PostgresDBConn and SQLiteDBConn.

PostgresDBConn provides high-performance data processing and support for concurrent operations, essential in collaborative and large-scale research settings. In contrast, SQLiteDBConn emphasizes simplicity, minimal setup, and ease of data sharing in single-user environments. This class is particularly suitable for scenarios that require a portable, file-based database system with a lower overhead.
During the importation process, the system automatically generates three distinct kinds of tables: static data tables that store information about ship characteristics, dynamic data tables that store ship movements, and aggregated data tables that compile an aggregate of static vessel reports for each unique Maritime Mobile Service Identity (MMSI) number. These tables provide a structured way to access detailed vessels' profiles, movements, and aggregated data points.


\subsubsection{Data Transformation}

The data values and trajectories can be transformed into other types suitable for user-defined queries and applications. For example, the system can convert datetime objects into string representations suitable for temporal queries. It also enables users to perform trajectory segmentation based on distance, velocity, and time between consecutive AIS messages from the same MMSI. The initial step of trajectory segmentation involves the representation of vessel trajectories as time series data, where each trajectory is defined by coordinates \((x, y)\) and time \(t\). The subsequent analysis calculates the speed \(v\) and distance \(m\) between consecutive points in the trajectory. The purpose of this functionality is to distinguish different voyages of a single vessel. This method uses thresholding on the calculated speed and distance to segment the trajectory data. Specifically, if \(v \leq V_{\text{max}}\) and \(\Delta m \leq M_{\text{max}}\), where \(V_{\text{max}}\) and \(M_{\text{max}}\) are predefined maximum values for speed and distance changes, a segment is considered for further analysis regardless of whether it is noisy or not.
AISdb facilitates the user in computing network graphs of vessel movements within different geographical zones. The polygon of each zone\footnote{A geographical area.} will be used as a network node, with graph edges represented by movements between zones. This enables users to perform network analyses for constructing maritime traffic networks.

\subsubsection{Noise Removal}

\begin{figure*}[!b]
    \includegraphics[width=.95\linewidth]{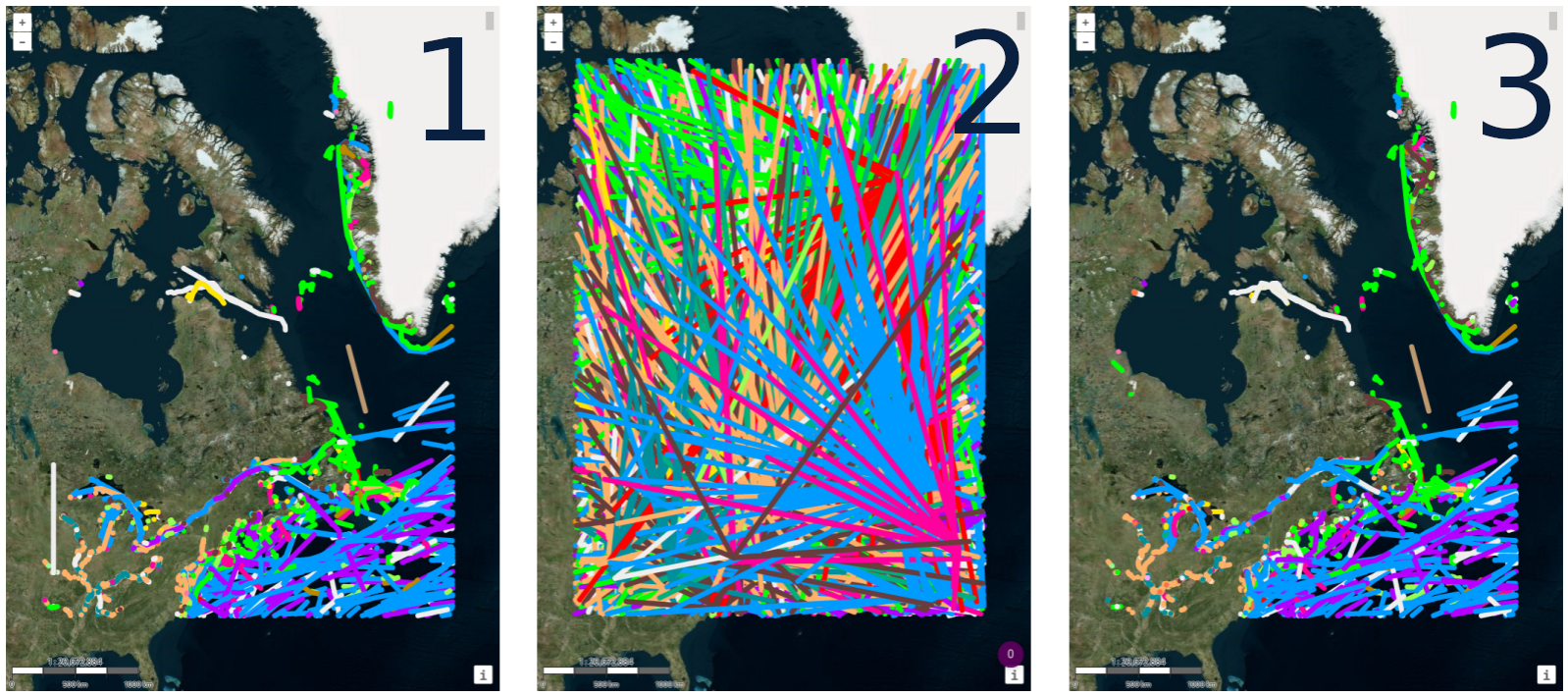}
    \caption{The image demonstrates the stages of applying the noise removal technique on vessel trajectory data from the North Atlantic Ocean: \textbf{(1)} Initial Data shows raw trajectories with inherent noise and variability; \textbf{(2)} Noise Dominated emphasizes the proliferation of noisy data points, creating a highly cluttered visual on top of \textit{Step (1)}; \textbf{(3)} Post-Processing reveals the cleaned trajectories after noise reduction, significantly improving the clarity and interpretability of vessel movements.}
    \label{fig:noise}
\end{figure*}

AISdb includes a noise removal technique for processing vessel trajectory data, which is essential for enhancing the accuracy of maritime monitoring systems.  
The function \(S\) calculates a score that indicates the likelihood of two consecutive points being part of the same trajectory as follows:
\begin{equation}
    S = 
    \begin{cases} 
        \frac{m}{\Delta m \times \Delta t} & \text{if } v \leq V_{\text{max}} \text{ and } \Delta m \leq M_{\text{max}}, \\
        -1 & \text{otherwise}.
    \end{cases}
\end{equation}
\noindent
where $m$ is the distance between points, $v$ is the velocity of the vessel and $\Delta t$ is the time taken between the two points.
The trajectory is reconstructed by concatenating segments with the highest scores, thereby filtering out segments considered as noise. This approach effectively cleans the noise trajectory data within the dataset and significantly improves the reliability of the tracking system used in maritime surveillance and traffic management. We exemplify the process using real and random data in Figure~\ref{fig:noise}.

\subsubsection{Data Filtering}

The data filtering module facilitates users in performing optimized searches for retrieving data based on various parameters, including specific time-frames, geographical bounding boxes, and vessel identifiers. Retrieving data within specified time frames facilitates temporal analysis of vessel movements.
Geographic bounding box query functionality is useful for regional traffic analysis. Additionally, the module allows users to filter results based on valid, invalid, or both types of MMSI numbers, providing different degrees of data reliability relevant to user-specific research needs.
The AISdb environment offers a callback module to post-process tracks during querying time. This module's functions are developed in Rust to connect with AISdb's Python functions, reducing computational complexity and allowing parallel computation; Python functions encapsulate these tools, making it easy to interact with AISdb.

\begin{figure}[!b] 
\begin{lstlisting}[language=Python]
from aisdb.database import sqlfcn_callbacks
from datetime import datetime, timedelta
from aisdb.track_gen import TrackGen
import aisdb

# Connection to a file-based database
with aisdb.SQLiteDBConn() as dbconn:
    now = datetime.now()
    
    # Query within the specified time range
    q = aisdb.DBQuery(
        dbconn=dbconn, dbpath='./sqlite-ais.db',
        callback=sqlfcn_callbacks.in_timerange,
        start=now - timedelta(hours=20),
        end=now, # 20-hour interval
    )

    # Generate data from the query, optionally returning a minimalistic trajectory after using a trajectory simplification algorithm
    rows = TrackGen(q.gen_qry(), decimate=False)

    # Split trajectories by time gaps, such as 24 hours without AIS message transmission
    rows = aisdb.split_timedelta(rows,
        timedelta(hours=24)
    )

    # Segmenting the data using a great circle distance filter based on distance and speed
    rows = aisdb.encode_greatcircledistance(rows,
        distance_threshold=200000,  # in meters
        speed_threshold=50          # in knots
    )

    # Interpolate time for uniform sampling
    tracks = aisdb.interp_time(rows,
        step=timedelta(minutes=5)
    )

    # Visualize the processed tracks using the AISdb web interface on the local browser
    aisdb.web_interface.visualize(
            open_browser=True,
            visualearth=True,
            tracks=tracks,
    )
\end{lstlisting}
\captionof{figure}{AIS data query, trajectory interpolation, and visualization Python script example using AISdb. The first step sets up a connection to an SQLite database and queries the last 20 hours of data contained in the database. The second step returns trajectories from the database that are split if time gaps are larger than 24 hours. The third step uses a great circle distance filter with a 200 km distance and 50 knots speed threshold to segment the trajectories further. In a fourth step, each trajectory is now interpolated with a time step of 5 minutes. Finally, the trajectories are plotted in a web interface.}
\label{ex_code}
\end{figure}

\subsubsection{Trajectory Interpolation}

Geospatial analyses rely on the GIS module, which provides various functions for defining and interacting with geographic areas. This module also includes features for distance calculations, time conversions, and spatial queries. Similarly, the interpolation module supports temporal data interpolation, crucial for creating consistent time-series datasets. AISdb facilitates spatial interpolation based on uniform temporal and equidistant intervals. To illustrate how to use these capabilities together, we provide a code example for querying, filtering, and further processing in Figure~\ref{ex_code}.

\subsubsection{Trajectory Visualization}

\begin{figure*}[ht]
  \centering
  \includegraphics[width=.95\textwidth]{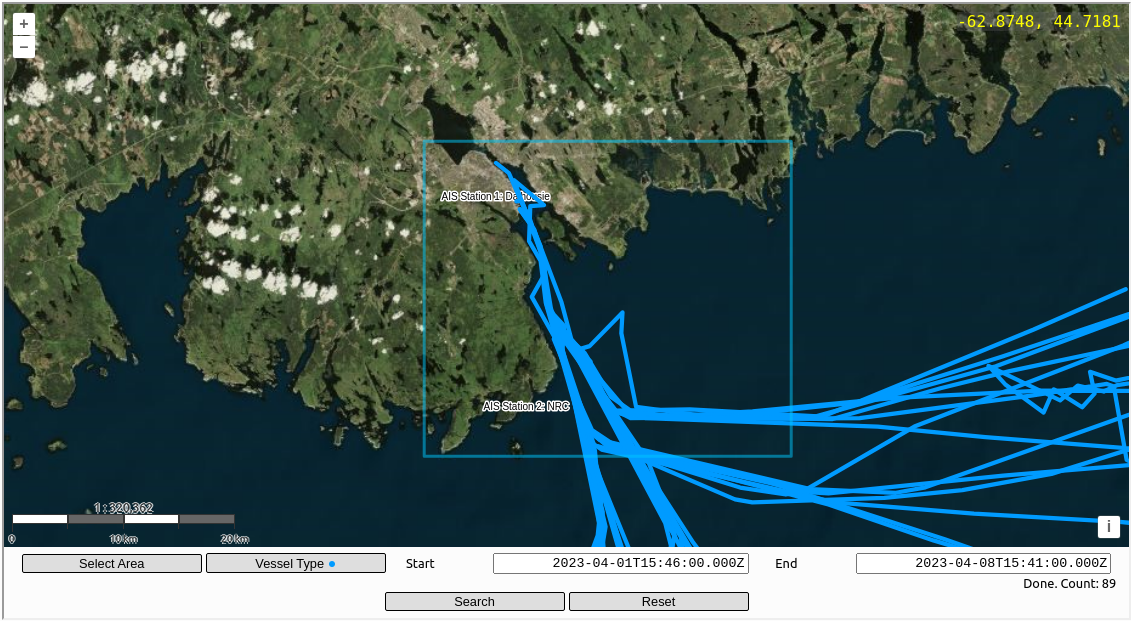}
  \caption{The web interface visualization showcases the diverse trajectories of vessels near our AIS Station antennas in Halifax at Dalhousie University and the National Research Council (NRC) Sandy Cove site, captured through AISdb’s tracking module and station. Vessel types are represented by different colors, and each line represents the ship's trajectory. The visualization illustrates how users can track vessels and query data using the web interface by selecting a region, range of dates, and vessel type.}
  \label{fig:example}
\end{figure*}

The AISdb Web Interface module facilitates interactive visualization and is deployed with WebAssembly to enable real-time processing and enhance the accessibility and comprehension of complex maritime datasets. Users can interact with the map, exploring the course of the trajectories, which have tools for filtering tracks by vessel type and identifier. In another module, AISdb introduces capabilities for estimating vessel-specific metrics, contributing to studies on fuel efficiency and environmental impact. This makes AISdb a comprehensive platform for analyzing maritime traffic and assessing its impact on the marine environment, serving the needs of researchers, policymakers, and industry professionals.
An in-browser visualization example is in Figure~\ref{fig:example}.

\subsubsection{Data Integration}

AISdb allows the integration of ocean and geographical features with maritime data, providing users with critical contextual insights. Integrating heterogeneous data sources is pivotal for numerous applications in the maritime domain, such as maritime navigation and security, marine traffic management, and environmental monitoring and protection. This integration is facilitated through specialized modules designed to access relevant data. As part of the AISdb tool, a dedicated module allows users to access and download free global bathymetric datasets\footnote{https://www.gebco.net}, enhancing the seamless integration of GeoTIFF-format bathymetry data with AIS data within AISdb. This functionality is essential for conducting depth analyses beneath vessel tracks, which can help users gain a deeper understanding of underwater topography concerning maritime navigation. Furthermore, the raster-based environmental data module provides tools for loading and querying environmental datasets, such as temperature or salinity maps. These tools are critical for enhancing the spatial analysis of maritime environments, allowing users to integrate and analyze geographic data layers effectively.

\subsubsection{Feature Extraction}

The AIS data for many vessels often lacks static information on the AIS messages sent by vessels, which contain important vessel characteristics and voyage information. To address this issue, AISdb enables users to access detailed vessel information and historical data from MarineTraffic. This enhances the AIS data with comprehensive details about vessel specifications and past activities for analysis.

AISdb has a module that calculates distances from geographic features like shorelines and ports. This module includes functions for calculating distances to and from these locations. These calculations are important for assessing potential risks, planning safer routes, ensuring navigational safety, and conducting environmental impact assessments.

\subsubsection{Data Exportation}

AISdb offers various export options, including the ability to export queried data (e.g., based on time frames and geographical areas) into CSV files. For those managing their own data stream or interested in conducting further analysis using Python tools (e.g., for machine learning applications), there's an option to securely store the data in a database and seamlessly integrate it into data processing pipelines.

\section{Scientific Impact}
\label{sec:sci}

AIS data has been crucial in various studies, including traffic analysis, environmental monitoring, safety improvements, and risk assessments. The AISdb, utilized at Dalhousie University, not only aggregates large-scale datasets but also finds application in analyzing smaller, region-specific datasets tailored for targeted research inquiries.

One example of AISdb usage is the work of Song et al. (2024)~\cite{song2024gravity}, in which the authors use the package to create graphs of marine mobility clustered at ports. These graphs are input to a machine learning workflow to identify the invasive species' spread risks through global transportation. Their study applies a physics-inspired gravity model incorporating transformers for maritime shipping traffic forecasting and classification of potential shipping directions. The data is then used on a two-stage model process involving machine learning classification to identify potential shipping flows followed by gravity-based flow estimation. Their technique achieves 89\% accuracy in trajectory segmentation and 84.8\% in vessel flow prediction; this approach marks a significant improvement over traditional models, providing valuable insights for managing the risks associated with non-indigenous species invasion and aiding policymakers and conservationists in strategic planning and mitigation efforts.

Another example of AIS data analysis is presented in the work by McWhinnie et al. (2021)~\cite{McWhinnie2021SalishSea}, in which the authors utilized AISdb paired with GIS technologies to process and visualize vessel traffic data in the Salish Sea. The researchers generated over 10 million linear segments from AIS positional data, enriching them with vessel type and temporal attributes. This data was aggregated into a $1km \times 1km$ grid for visual analysis, allowing for detailed spatio-temporal exploration of vessel distributions across critical habitats. The study revealed seasonal variability in vessel traffic, particularly the dominance of ferries in summer months and the prevalence of cargo ships throughout the year, especially in Boundary Pass. Their study suggests targeted vessel management measures to mitigate environmental impacts, emphasizing the importance of seasonal and vessel-type-specific regulations to protect the endangered Southern Resident Killer Whale population. The findings contribute valuable insights for marine spatial planning, species recovery measures, and conservation strategies.

An ecological application of the AISdb was to identify and inform Canadian marine invasive species managers of vessel traffic that facilitates the secondary (local) spread of non-indigenous species. The project's overarching objective was to identify biogeographic boundaries (or transition zones) that limit or contain the natural dispersal and spread of marine invasive species in temperate, subarctic, and arctic marine ecosystems~\cite{Krumhansl2023Permeability}. These barriers are compromised by ballast and hull biofouling vectors mediating marine species transfer between otherwise disjunct ecoregions. Vessel network analysis was employed to identify geographic pathways of anthropogenic vectors that could compromise natural barriers to dispersal. Our results identified natural barriers, vectors, and pathways in support of the development of targeted management measures for the prevention of the anthropogenic spread of non-indigenous species~\cite{Lyons2023Identifying}.

In the work by Hu et al. (2016)~\cite{Baifan2016fishing}, the authors utilized Conditional Random Fields (CRFs) to identify fishing activities from Automatic Identification System (AIS) data. CRFs effectively solve structured prediction problems and were used to model the conditional probability distributions that distinguish fishing from non-fishing activities based on AIS attributes like coordinates and speeds. The study involved extensive data preprocessing, including data cleaning, discretization, and feature selection with the aid of AISdb, followed by three experiments and two comparisons to evaluate the model's performance. The results demonstrated that the CRF model achieved an average accuracy of 88.7\%, outperforming other methods such as autoencoders and data mining approaches. This study highlights the potential of CRFs in providing accurate and efficient fishing activity detection, contributing valuable insights for sustainable fishery management and the enforcement of regulations against illegal, unreported, and unregulated (IUU) fishing activities

A final example relates to AIS trajectory forecasting presented in the work of Spadon et al. (2024)~\cite{spadon2024building}, which utilizes a deep autoencoder model combined with a phased framework approach to predict long-term vessel trajectories. They fuse spatiotemporal features from AIS messages with probabilistic features engineered from historical AIS data, transforming the forecasting challenge into a trajectory reconstruction problem. Their study innovatively applies this method to predict the next 12 hours of vessel trajectories using 1 to 3 hours of AIS data as input. The model achieves an F1-Score of approximately 85\% and 75\% for vessel route and destination prediction, respectively, and an R2 Score of over 98\% for various feature combinations. This technique demonstrates a 25\% improvement in precision over state-of-the-art methods, with average and median errors of 11 km and 6 km, respectively. The model is deployed as part of the \textit{smartWhales} initiative, contributing to marine safety by predicting vessel routes and preventing vessel-whale collisions in the \textit{Gulf of St. Lawrence}, providing significant advancements in trajectory forecasting and aiding policymakers and conservationists in strategic planning and mitigation efforts.

As demonstrated, AISdb has played a significant role in advancing maritime research, particularly at Dalhousie University. These previous studies demonstrate the potential of AISdb in driving improvements in maritime safety, environmental protection, and species preservation. This reinforces the claim that AISdb can be broadly used by the research community interested in enhancing marine research and safeguarding our oceans for future generations.

\section{Conclusions}
\label{sec:conclusions}

AISdb has become an essential tool for maritime research, serving as a fundamental resource for numerous studies focused on improving maritime safety, promoting environmental sustainability, and gaining insight into vessel behavior. This paper demonstrates that AISdb is utilized for various purposes, including detailed traffic analysis, environmental monitoring, and predictive vessel trajectory forecasting. Utilizing AISdb and AIS data, researchers have expanded the scope of predictive modeling and risk assessment, resulting in contributions to preserving marine ecosystems and enhancing safety in maritime navigation.

The scope and applicability of AISdb in marine research are set to expand in the future. Plans are in place to continuously update and broaden the database with more diverse datasets and analytical tools to cater to a wider range of research inquiries and their researchers. These enhancements will deepen the quality of maritime studies and attract a larger community of researchers eager to explore the vast potential of AIS data. Additionally, AISdb intends to collaborate with maritime authorities and environmental organizations to bring real-world applications of its research findings. Such collaborations will directly contribute to maritime policy-making, conservation efforts, and the global endeavor toward safer and more sustainable maritime operations.

The project and funding source of AISdb entitled ``AISviz: Making Vessels Tracking Data and Maps Available to Everyone'', aims to enhance national capacity to access, process, and visualize historical and real-time Automated Identification Systems vessel traffic data. This will enable monitoring and prediction of the impact of commercial and leisure vessels in current and future marine protected areas and other effective area-based conservation measures. As part of this initiative, we aim to expand the functionalities and provide active support until 2026 and possibly beyond.


\section*{Acknowledgments}
\noindent%
This research was supported by the Institute for Big Data Analytics (IBDA) and the Ocean Frontier Institute (OFI) at Dalhousie University in Halifax, NS, Canada. Additionally, it received further funding from the Canada First Research Excellence Fund (CFREF) and the Canadian Foundation for Innovation's MERIDIAN Cyberinfrastructure. The software development was partially funded by the Department of Fisheries and Oceans-funded project ``AISviz: Making Vessels Tracking Data and Maps Available to Everyone''.


\section*{Data availability}
\noindent%
Directions on how to use AISdb with open-source datasets are provided with examples at \href{https://aisviz.gitbook.io/tutorials/}{https://aisviz.gitbook.io/tutorials/}.

\section*{Generative AI Acknowledgement}
\noindent%
We have utilized Generative Artificial Intelligence (GenAI) models, specifically Large Language Models (LLMs), to assist in content revision. These tools have played a supportive role, by helping to rethink and rewrite sentences for improved clarity and coherence. It is important to note that all content presented has undergone extensive revision and rewriting by the authors. The intellectual contribution, final wording, and overall responsibility for the content rest solely with the human authors, reassuring this work's integrity.


\bibliographystyle{unsrtnat}
\bibliography{cas-refs}

\begin{thebibliography}{44}
\providecommand{\natexlab}[1]{#1}
\providecommand{\url}[1]{\texttt{#1}}
\expandafter\ifx\csname urlstyle\endcsname\relax
  \providecommand{\doi}[1]{doi: #1}\else
  \providecommand{\doi}{doi: \begingroup \urlstyle{rm}\Url}\fi

\bibitem[Organization({\natexlab{a}})]{AIS}
International~Maritime Organization.
\newblock Ais transponders.
\newblock \url{https://www.imo.org/en/OurWork/Safety/Pages/AIS.aspx}, {\natexlab{a}}.
\newblock Accessed: 17th October 2022.

\bibitem[Organization({\natexlab{b}})]{IMO}
International~Maritime Organization.
\newblock Brief history of imo.
\newblock \url{https://www.imo.org/en/About/HistoryOfIMO/Pages/Default.aspx}, {\natexlab{b}}.
\newblock Accessed: 17th October 2022.

\bibitem[Peng et~al.(2022)Peng, Wang, Tong, Zhang, Zou, and Tan]{peng2022establishment}
Zhongbo Peng, Lumeng Wang, Liang Tong, Chunyu Zhang, Han Zou, and Jianping Tan.
\newblock Establishment of inland ship air pollution emission inventory based on power method correction model.
\newblock \emph{Sustainability}, 14\penalty0 (18):\penalty0 11188, 2022.

\bibitem[Yang et~al.(2019)Yang, Wu, Wang, Jia, and Li]{yang2019big}
Dong Yang, Lingxiao Wu, Shuaian Wang, Haiying Jia, and Kevin~X Li.
\newblock How big data enriches maritime research--a critical review of automatic identification system (ais) data applications.
\newblock \emph{Transport Reviews}, 39\penalty0 (6):\penalty0 755--773, 2019.

\bibitem[Fournier et~al.(2018)Fournier, Casey~Hilliard, Rezaee, and Pelot]{fournier2018past}
M{\'e}lanie Fournier, R~Casey~Hilliard, Sara Rezaee, and Ronald Pelot.
\newblock Past, present, and future of the satellite-based automatic identification system: Areas of applications (2004--2016).
\newblock \emph{WMU journal of maritime affairs}, 17:\penalty0 311--345, 2018.

\bibitem[Goerlandt and Kujala(2011)]{Goerlandt2011:ShipCollision}
Floris Goerlandt and Pentti Kujala.
\newblock Traffic simulation based ship collision probability modeling.
\newblock \emph{Reliability Engineering \& System Safety}, 96\penalty0 (1):\penalty0 91--107, 2011.
\newblock ISSN 0951-8320.
\newblock \doi{10.1016/j.ress.2010.09.003}.
\newblock Special Issue on Safecomp 2008.

\bibitem[Chen et~al.(2019{\natexlab{a}})Chen, Huang, Mou, and van Gelder]{Chen2019:ShipCollision}
Pengfei Chen, Yamin Huang, Junmin Mou, and P.H.A.J.M. van Gelder.
\newblock Probabilistic risk analysis for ship-ship collision: State-of-the-art.
\newblock \emph{Safety Science}, 117:\penalty0 108--122, 2019{\natexlab{a}}.
\newblock ISSN 0925-7535.
\newblock \doi{10.1016/j.ssci.2019.04.014}.

\bibitem[Harati-Mokhtari et~al.(2007)Harati-Mokhtari, Wall, Brooks, and Wang]{harati2007automatic}
Abbas Harati-Mokhtari, Alan Wall, Philip Brooks, and Jin Wang.
\newblock Automatic identification system (ais): data reliability and human error implications.
\newblock \emph{The Journal of Navigation}, 60\penalty0 (3):\penalty0 373--389, 2007.

\bibitem[Millefiori et~al.(2021)Millefiori, Braca, Zissis, Spiliopoulos, Marano, Willett, and Carniel]{millefiori2021covid}
Leonardo~M Millefiori, Paolo Braca, Dimitris Zissis, Giannis Spiliopoulos, Stefano Marano, Peter~K Willett, and Sandro Carniel.
\newblock Covid-19 impact on global maritime mobility.
\newblock \emph{Scientific reports}, 11\penalty0 (1):\penalty0 1--16, 2021.

\bibitem[Fu et~al.(2017)Fu, Wang, Liu, Hu, and Zhang]{fu2017finding}
Peiguo Fu, Haozhou Wang, Kuien Liu, Xiaohui Hu, and Hui Zhang.
\newblock Finding abnormal vessel trajectories using feature learning.
\newblock \emph{IEEE Access}, 5:\penalty0 7898--7909, 2017.

\bibitem[Li et~al.(2018)Li, Liu, Wu, Yang, Liu, and Xiong]{li2018spatio}
Huanhuan Li, Jingxian Liu, Kefeng Wu, Zaili Yang, Ryan~Wen Liu, and Naixue Xiong.
\newblock Spatio-temporal vessel trajectory clustering based on data mapping and density.
\newblock \emph{IEEE Access}, 6:\penalty0 58939--58954, 2018.

\bibitem[Haranwala et~al.(2023)Haranwala, Spadon, Renso, and Soares]{haranwala2023data}
Yaksh~J. Haranwala, Gabriel Spadon, Chiara Renso, and Amilcar Soares.
\newblock A data augmentation algorithm for trajectory data.
\newblock In \emph{1st ACM SIGSPATIAL International Workshop on Methods for Enriched Mobility Data: Emerging issues and Ethical perspectives 2023 (EMODE '23)}, page~5, New York, NY, USA, 2023. ACM, New York, NY, USA.
\newblock \doi{10.1145/3615885.3628008}.

\bibitem[Mustafa et~al.(2021)Mustafa, Barrus, Leal, and Gruenwald]{mustafa2021gtraclus}
Hamza Mustafa, Clark Barrus, Eleazar Leal, and Le~Gruenwald.
\newblock Gtraclus: A local trajectory clustering algorithm for gpus.
\newblock In \emph{2021 IEEE 37th International Conference on Data Engineering Workshops (ICDEW)}, pages 30--35. IEEE, 2021.

\bibitem[Rong et~al.(2020)Rong, Teixeira, and Soares]{rong2020data}
H~Rong, AP~Teixeira, and C~Guedes Soares.
\newblock Data mining approach to shipping route characterization and anomaly detection based on ais data.
\newblock \emph{Ocean Engineering}, 198:\penalty0 106936, 2020.

\bibitem[Nguyen et~al.(2018)Nguyen, Vadaine, Hajduch, Garello, and Fablet]{nguyen2018multi}
Duong Nguyen, Rodolphe Vadaine, Guillaume Hajduch, Ren{\'e} Garello, and Ronan Fablet.
\newblock A multi-task deep learning architecture for maritime surveillance using ais data streams.
\newblock In \emph{2018 IEEE 5th International Conference on Data Science and Advanced Analytics (DSAA)}, pages 331--340. IEEE, 2018.

\bibitem[Patmanidis et~al.(2016)Patmanidis, Voulgaris, Sarri, Papavassilopoulos, and Papavasileiou]{patmanidis2016maritime}
Spyridon Patmanidis, Iasonas Voulgaris, Elena Sarri, George Papavassilopoulos, and George Papavasileiou.
\newblock Maritime surveillance, vessel route estimation and alerts using ais data.
\newblock In \emph{2016 24th Mediterranean Conference on Control and Automation (MED)}, pages 809--813. IEEE, 2016.

\bibitem[Zhang et~al.(2018)Zhang, Peng, Liu, and Liu]{zhang2018ais}
Ya-lun Zhang, Peng-fei Peng, Jian-shu Liu, and Shu-kan Liu.
\newblock Ais data oriented ships' trajectory mining and forecasting based on trajectory delimiter.
\newblock In \emph{2018 10th International Conference on Intelligent Human-Machine Systems and Cybernetics (IHMSC)}, volume~1, pages 269--273. IEEE, 2018.

\bibitem[{\"U}ney et~al.(2019){\"U}ney, Millefiori, and Braca]{uney2019data}
Murat {\"U}ney, Leonardo~M Millefiori, and Paolo Braca.
\newblock Data driven vessel trajectory forecasting using stochastic generative models.
\newblock In \emph{IEEE International Conference on Acoustics, Speech and Signal Processing (ICASSP)}, pages 8459--8463. IEEE, 2019.

\bibitem[Spadon et~al.(2022)Spadon, Ferreira, Soares, and Matwin]{spadon2022unfolding}
Gabriel Spadon, Martha~D Ferreira, Amilcar Soares, and Stan Matwin.
\newblock Unfolding ais transmission behavior for vessel movement modeling on noisy data leveraging machine learning.
\newblock \emph{IEEE Access}, 2022.

\bibitem[Newaliya and Singh(2021)]{newaliya2021review}
Nitin Newaliya and Yudhvir Singh.
\newblock A review of maritime spatio-temporal data analytics.
\newblock In \emph{2021 International Conference on Computational Performance Evaluation (ComPE)}, pages 219--226. IEEE, 2021.

\bibitem[Le~Guillarme and Lerouvreur(2013)]{le2013unsupervised}
Nicolas Le~Guillarme and Xavier Lerouvreur.
\newblock Unsupervised extraction of knowledge from s-ais data for maritime situational awareness.
\newblock In \emph{Proceedings of the 16th International Conference on Information Fusion}, pages 2025--2032. IEEE, 2013.

\bibitem[d'Afflisio et~al.(2018)d'Afflisio, Braca, Millefiori, and Willett]{d2018maritime}
Enrica d'Afflisio, Paolo Braca, Leonardo~M Millefiori, and Peter Willett.
\newblock Maritime anomaly detection based on mean-reverting stochastic processes applied to a real-world scenario.
\newblock In \emph{2018 21st International Conference on Information Fusion (FUSION)}, pages 1171--1177. IEEE, 2018.

\bibitem[Forti et~al.(2019{\natexlab{a}})Forti, Millefiori, and Braca]{forti2019unsupervised}
Nicola Forti, Leonardo~M Millefiori, and Paolo Braca.
\newblock Unsupervised extraction of maritime patterns of life from automatic identification system data.
\newblock In \emph{OCEANS 2019-Marseille}, pages 1--5. IEEE, 2019{\natexlab{a}}.

\bibitem[Campbell et~al.(2022)Campbell, Isenor, and Ferreira]{campbell2022detection}
Jessica~NA Campbell, Anthony~W Isenor, and Martha~Dais Ferreira.
\newblock Detection of invalid ais messages using machine learning techniques.
\newblock \emph{Procedia Computer Science}, 205:\penalty0 229--238, 2022.

\bibitem[Forti et~al.(2019{\natexlab{b}})Forti, Millefiori, Braca, and Willett]{forti2019anomaly}
Nicola Forti, Leonardo~M Millefiori, Paolo Braca, and Peter Willett.
\newblock Anomaly detection and tracking based on mean--reverting processes with unknown parameters.
\newblock In \emph{ICASSP 2019-2019 IEEE International Conference on Acoustics, Speech and Signal Processing (ICASSP)}, pages 8449--8453. IEEE, 2019{\natexlab{b}}.

\bibitem[Pichegru et~al.(2022)Pichegru, Vibert, Thiebault, Charrier, Stander, Ludynia, Lewis, Carpenter-Kling, and McInnes]{pichegru2022maritime}
Lorien Pichegru, La{\"e}titia Vibert, Andr{\'e}a Thiebault, Isabelle Charrier, Nicky Stander, Katta Ludynia, Melissa Lewis, Tegan Carpenter-Kling, and Alistair McInnes.
\newblock Maritime traffic trends around the southern tip of africa--did marine noise pollution contribute to the local penguins' collapse?
\newblock \emph{Science of The Total Environment}, 849:\penalty0 157878, 2022.

\bibitem[Faghih-Roohi et~al.(2014)Faghih-Roohi, Xie, and Ng]{faghih2014accident}
Shahrzad Faghih-Roohi, Min Xie, and Kien~Ming Ng.
\newblock Accident risk assessment in marine transportation via markov modeling and markov chain monte carlo simulation.
\newblock \emph{Ocean engineering}, 91:\penalty0 363--370, 2014.

\bibitem[AbuAlhaol et~al.(2018)AbuAlhaol, Falcon, Abielmona, and Petriu]{abualhaol2018mining}
Ibrahim AbuAlhaol, Rafael Falcon, Rami Abielmona, and Emil Petriu.
\newblock Mining port congestion indicators from big ais data.
\newblock In \emph{2018 International Joint Conference on Neural Networks (IJCNN)}, pages 1--8. IEEE, 2018.

\bibitem[Varlamis et~al.(2018)Varlamis, Tserpes, and Sardianos]{varlamis2018detecting}
Iraklis Varlamis, Konstantinos Tserpes, and Christos Sardianos.
\newblock Detecting search and rescue missions from ais data.
\newblock In \emph{2018 IEEE 34th international conference on data engineering workshops (ICDEW)}, pages 60--65. IEEE, 2018.

\bibitem[Zhou et~al.(2020)Zhou, Chen, and Zhang]{zhou2020performance}
Fan Zhou, Hua Chen, and Peng Zhang.
\newblock Performance evaluation of maritime search and rescue missions using automatic identification system data.
\newblock \emph{The Journal of Navigation}, 73\penalty0 (6):\penalty0 1237--1246, 2020.

\bibitem[Chen et~al.(2019{\natexlab{b}})Chen, Sun, Liang, Han, Su, et~al.]{chen2019study}
Weijiong Chen, Yuejiao Sun, Yu~Liang, Zewei Han, Junfang Su, et~al.
\newblock Study on buoys for trajectory prediction simulation of maritime drifting objects.
\newblock \emph{Academic Journal of Engineering and Technology Science}, 2\penalty0 (2), 2019{\natexlab{b}}.

\bibitem[Ferreira et~al.(2022)Ferreira, Spadon, Soares, and Matwin]{ferreira2022semi}
Martha~Dais Ferreira, Gabriel Spadon, Amilcar Soares, and Stan Matwin.
\newblock A semi-supervised methodology for fishing activity detection using the geometry behind the trajectory of multiple vessels.
\newblock \emph{Sensors}, 22\penalty0 (16):\penalty0 6063, 2022.

\bibitem[Li et~al.(2023)Li, Ren, and Li]{DBLP:journals/softx/LiRL23}
Ye~Li, Hongxiang Ren, and Haijiang Li.
\newblock Pyvt: {A} toolkit for preprocessing and analysis of vessel spatio-temporal trajectories.
\newblock \emph{SoftwareX}, 21:\penalty0 101316, 2023.
\newblock \doi{10.1016/J.SOFTX.2023.101316}.
\newblock URL \url{https://doi.org/10.1016/j.softx.2023.101316}.

\bibitem[Sanches(2019)]{arina2019}
Arina De Jesus Amador~Monteiro Sanches.
\newblock Uma arquitetura e implementação do módulo de pré-processamento para biblioteca pymove.
\newblock Bachelor's thesis, Universidade Federal Do Ceará, 2019.

\bibitem[Pappalardo et~al.(2022)Pappalardo, Simini, Barlacchi, and Pellungrini]{DBLP:journals/jstatsoft/PappalardoSBP22}
Luca Pappalardo, Filippo Simini, Gianni Barlacchi, and Roberto Pellungrini.
\newblock scikit-mobility: {A} \emph{Python} library for the analysis, generation, and risk assessment of mobility data.
\newblock 103\penalty0 (4), 2022.
\newblock \doi{10.18637/JSS.V103.I04}.
\newblock URL \url{https://doi.org/10.18637/jss.v103.i04}.

\bibitem[Haidri et~al.(2022)Haidri, Haranwala, Bogorny, Renso, da~Fonseca, and Soares]{DBLP:journals/softx/HaidriHBRFS22}
Salman Haidri, Yaksh~J. Haranwala, Vania Bogorny, Chiara Renso, Vinicius~Prado da~Fonseca, and Am{\'{\i}}lcar Soares.
\newblock {PTRAIL} - {A} python package for parallel trajectory data preprocessing.
\newblock \emph{SoftwareX}, 19:\penalty0 101176, 2022.
\newblock \doi{10.1016/J.SOFTX.2022.101176}.
\newblock URL \url{https://doi.org/10.1016/j.softx.2022.101176}.

\bibitem[Kontopoulos et~al.(2023)Kontopoulos, Makris, and Tserpes]{DBLP:journals/softx/KontopoulosMT23}
Ioannis Kontopoulos, Antonios Makris, and Konstantinos Tserpes.
\newblock Traclets: {A} trajectory representation and classification library.
\newblock \emph{SoftwareX}, 21:\penalty0 101306, 2023.
\newblock \doi{10.1016/J.SOFTX.2023.101306}.
\newblock URL \url{https://doi.org/10.1016/j.softx.2023.101306}.

\bibitem[Tang et~al.(2021)Tang, Wang, Zhao, Tang, Yan, and Xiao]{TANG2021109041}
Chunhua Tang, Han Wang, Jiahuan Zhao, Yuanqing Tang, Huaran Yan, and Yingjie Xiao.
\newblock A method for compressing ais trajectory data based on the adaptive-threshold douglas-peucker algorithm.
\newblock \emph{Ocean Engineering}, 232:\penalty0 109041, 2021.
\newblock ISSN 0029-8018.
\newblock \doi{https://doi.org/10.1016/j.oceaneng.2021.109041}.
\newblock URL \url{https://www.sciencedirect.com/science/article/pii/S0029801821004765}.

\bibitem[Song et~al.(2024)Song, Spadon, Pelot, Matwin, and Soares]{song2024gravity}
Ruixin Song, Gabriel Spadon, Ronald Pelot, Stan Matwin, and Amilcar Soares.
\newblock Gravity-informed deep learning framework for predicting ship traffic flow and invasion risk of non-indigenous species via ballast water discharge.
\newblock \emph{arXiv}, 2024.

\bibitem[McWhinnie et~al.(2021)McWhinnie, O'Hara, Hilliard, {Le Baron}, Smallshaw, Pelot, and Canessa]{McWhinnie2021SalishSea}
Lauren~H. McWhinnie, Patrick~D. O'Hara, Casey Hilliard, Nicole {Le Baron}, Leh Smallshaw, Ronald Pelot, and Rosaline Canessa.
\newblock Assessing vessel traffic in the salish sea using satellite ais: An important contribution for planning, management and conservation in southern resident killer whale critical habitat.
\newblock \emph{Ocean \& Coastal Management}, 200:\penalty0 105479, 2021.
\newblock ISSN 0964-5691.
\newblock \doi{https://doi.org/10.1016/j.ocecoaman.2020.105479}.

\bibitem[Krumhansl et~al.(2023)Krumhansl, Gentleman, Lee, Ramey-Balci, Goodwin, Wang, Lowen, Lyons, Therriault, and DiBacco]{Krumhansl2023Permeability}
Kira Krumhansl, Wendy Gentleman, Katherine Lee, Patricia Ramey-Balci, Jace Goodwin, Zeliang Wang, Ben Lowen, Devin Lyons, Thomas~W. Therriault, and Claudio DiBacco.
\newblock Permeability of coastal biogeographic barriers to marine larval dispersal on the east and west coasts of north america.
\newblock \emph{Global Ecology and Biogeography}, 32\penalty0 (6):\penalty0 945--961, 2023.
\newblock \doi{https://doi.org/10.1111/geb.13654}.

\bibitem[Lyons et~al.(2023)Lyons, Krumhansl, Smith, Therriault, Gentleman, Lowen, Ramey-Balci, Wang, and DiBacco]{Lyons2023Identifying}
D.~A. Lyons, K.~A. Krumhansl, M.~R. Smith, T.~W. Therriault, W.~C. Gentleman, J.~B. Lowen, P.~A. Ramey-Balci, Z.~Wang, and C.~DiBacco.
\newblock Identifying natural biogeographic barriers in support of aquatic invasive species management.
\newblock In \emph{International Conference on Marine Bioinvasions}, Baltimore, Maryland, May 2023.
\newblock Presented at the International Conference on Marine Bioinvasions, 15-19 May 2023.

\bibitem[Hu et~al.(2016)Hu, Jiang, de~Souza, Pelot, and Matwin]{Baifan2016fishing}
Baifan Hu, Xiang Jiang, Erico~N de~Souza, Ronald Pelot, and Stan Matwin.
\newblock Identifying fishing activities from ais data with conditional random fields.
\newblock In \emph{2016 Federated Conference on Computer Science and Information Systems (FedCSIS)}, pages 47--52, 2016.

\bibitem[Spadon et~al.(2024)Spadon, Kumar, Eden, van Berkel, Foster, Soares, Fablet, Matwin, and Pelot]{spadon2024building}
Gabriel Spadon, Jay Kumar, Derek Eden, Josh van Berkel, Tom Foster, Amilcar Soares, Ronan Fablet, Stan Matwin, and Ronald Pelot.
\newblock Probabilistic feature augmentation for ais-based multi-path long-term vessel trajectory forecasting.
\newblock \emph{arXiv}, 2024.

\end{thebibliography}

\balance

\end{document}